# Managing Distributed Software Development in the Virtual Astronomical Observatory

Janet D. Evans*[a], Raymond L. Plante[b], Nina Bonaventura[a,], Ivo Busko[c], Mark Cresitello-Dittmar[a], Raffaele D'Abrusco[a], Stephen Doe[a], Rick Ebert[d], Omar Laurino[a], Olga Pevunova[d], Brian Refsdal[†a], Brian Thomas[e]

[a]Smithsonian Astrophysical Observatory, 60 Garden St., Cambridge, MA, USA 02138; [b]University of Illinois at Urbana-Champaign, 601 E. John St., Champaign, IL USA 61820; [c]Space Telescope Science Institute, 3700 San Martin Dr., Baltimore, MD USA 21218; [d]Infrared Processing and Analysis Center, 770 South Wilson Ave., Pasadena, CA USA 91125; [e]National Optical Astronomy Observatory, 950 Cherry Ave., Tucson, MA USA 85719

## ABSTRACT

The U.S. Virtual Astronomical Observatory (VAO) is a product-driven organization that provides new scientific research capabilities to the astronomical community. Software development for the VAO follows a lightweight framework that guides development of science applications and infrastructure. Challenges to be overcome include distributed development teams, part-time efforts, and highly constrained schedules. We describe the process we followed to conquer these challenges while developing *Iris*, the VAO application for analysis of 1-D astronomical spectral energy distributions (SEDs). Iris was successfully built and released in less than a year with a team distributed across four institutions. The project followed existing International Virtual Observatory Alliance inter-operability standards for spectral data and contributed a SED library as a by-product of the project. We emphasize lessons learned that will be folded into future development efforts. In our experience, a well-defined process that provides guidelines to ensure the project is cohesive and stays on track is key to success. Internal product deliveries with a planned test and feedback loop are critical. Release candidates are measured against use cases established early in the process, and provide the opportunity to assess priorities and make course corrections during development. Also key is the participation of a stakeholder such as a lead scientist who manages the technical questions, advises on priorities, and is actively involved as a lead tester. Finally, frequent scheduled communications (for example a bi-weekly tele-conference) assure issues are resolved quickly and the team is working toward a common vision.

## 1. INTRODUCTION

Historically, astronomical science applications have been managed and developed at a single institution and often by a single individual. Due to the size and scope of projects, reduced budgets, and the ease at which the digital age has allowed for more reasonable communication, software development efforts are now often distributed; and not only at the project level but also at the single application level. The Iris[1] team has representatives from 5 institutions (SAO, STScI, IPAC, NOAO, and NCSA) well known in astronomy. Expertise ranges from knowledge of specific areas of astronomical data analysis (e.g., optical or x-ray), to expertise in data storage and access (e.g., NED, Hubble, Chandra); scientific knowledge of spectral data analysis, to experience with International Virtual Observatory Alliance[1] (IVOA) Data Model and access protocol standards. Many of the team members met for the first time at the start of the project and each team member has critical experience and talent to contribute to the project. No one individual could have successfully completed the project on his or her own given the time constraints, but as a team we met the development challenge and Iris product release.

---

*janet@cfa.harvard.edu; phone 1 617 495-7160; fax 1 617 496-1693; http://www.cfa.harvard.edu/
[†]Current address: SAY Media Inc., 180 Townsend St., San Francisco, CA, USA 94107
[1]http://www.ivoa.net/

Updated 1 March 2012

This paper is about how we got the job done within the Virtual Astronomical Observatory[2] (VAO), and the important elements to consider when working with a distributed team reaching broadly over the astronomical research community and with a tightly constrained schedule.

## 2. VAO ECOSYSTEM

The *VAO Ecosystem* refers to the integrated research environment that the VAO is working to provide for astronomers. It addresses all phases of the research process from data and knowledge discovery, through processing and analysis, and to data publishing. Central to the ecosystem is access to the vast variety of scientific data from on-line astronomical archives (both public and proprietary). In addition to access to existing data, the ecosystem encourages the publishing of small, highly processed data collections upon which the results in the published literature are based but which are typically unavailable. As an ecosystem, the implied architecture is one that is open, loosely-coupled and yet smoothly connected, bringing together VAO-provided tools with top-quality tools developed by the wider Virtual Observatory (VO) community. In particular, VAO science tools provide key end-user capabilities that make the environment usable but also pave the way for community-contributed tools. The VAO and community tools will rely on a persistent infrastructure (supporting publishing, discovery, identity management, portal integration, and tool development) with high reliability.

The VAO has a Standards and Infrastructure team providing the components, libraries, and templates that enable end-users the ability to craft their own VO-enabled applications for data access and integration. We also have a Science Applications team that is developing and supporting general-purpose science applications that provide feedback to the infrastructure components and identify where further development is needed. At the same time, we provide science capabilities and examples to the user community on how applications are built and work together with the available infrastructure and services. Iris is an example of a science application built to serve users while utilizing and providing feedback to infrastructure components built for interoperability.

## 3. IRIS

*Iris* is a downloadable graphical user interface (GUI) application that enables astronomers to build and analyze wide-band Spectral Energy Distributions (SEDs; Fig. 1). SED data may be loaded into Iris from a file on the user's local disk in ASCII, FITS, or VOTable format, imported directly from the NASA Extragalactic Database[3] (NED), or received from the VAO Data Discovery Tool[2] (DDT), or any other VO-enabled application, for analysis.

Iris consists of new software along with the integration of existing packages to form a new application for SED analysis (Fig. 2). The SED builder component reads in SEDs and optionally converts non-standard SED data files into a format supported by Iris. Specview[4], contributed by Space Telescope Science Institute (STScI), provides a GUI for reading, editing, and visualizing SEDs, as well as defining model expressions and setting initial model parameter values. Sherpa[5], contributed by the Smithsonian Astrophysical Observatory (SAO), provides a library of models, fit statistics, and optimization methods for analyzing SEDs. The underlying i/o library (SEDlib), also contributed by SAO, is written to the IVOA Spectrum Data Model standard. NED is a service provided by the Infrared Processing and Analysis Center (IPAC) at the California Institute for Technology for easy location of the data available for a given extragalactic astronomical source, including SEDs.

We built infrastructure that provides an API to all components and manages events as defined in the interface. Iris interoperates with other applications via a Simple Application Messaging Protocol[6] (SAMP) interface and sends and receives SED data through that connection. Another key component for interoperability built into Iris is the capability to provide user-defined plug-ins. An external user can write a module to manage their specific data format or do analysis on their data with a scripted method. The new module can be added simply to Iris by using a lightweight Java template that we provide and the addition of user specific code. Interoperability is a key component of Iris and we use it in three ways; (1) by reusing software from the astronomical community (e.g., Specview and Sherpa) in a tightly coupled application (implemented via SAMP) to enable new science analysis; (2) by providing extension capabilities so that any group can write their own module (e.g., read instrument specific SED format from an archive or do SED specific analysis); and (3) by implementing hooks to send and receive SED data from other tools with similar SAMP interfaces.

---

[2]http://www.usvao.org/science-tools-services/vao-tools-services-data-discovery-tool/

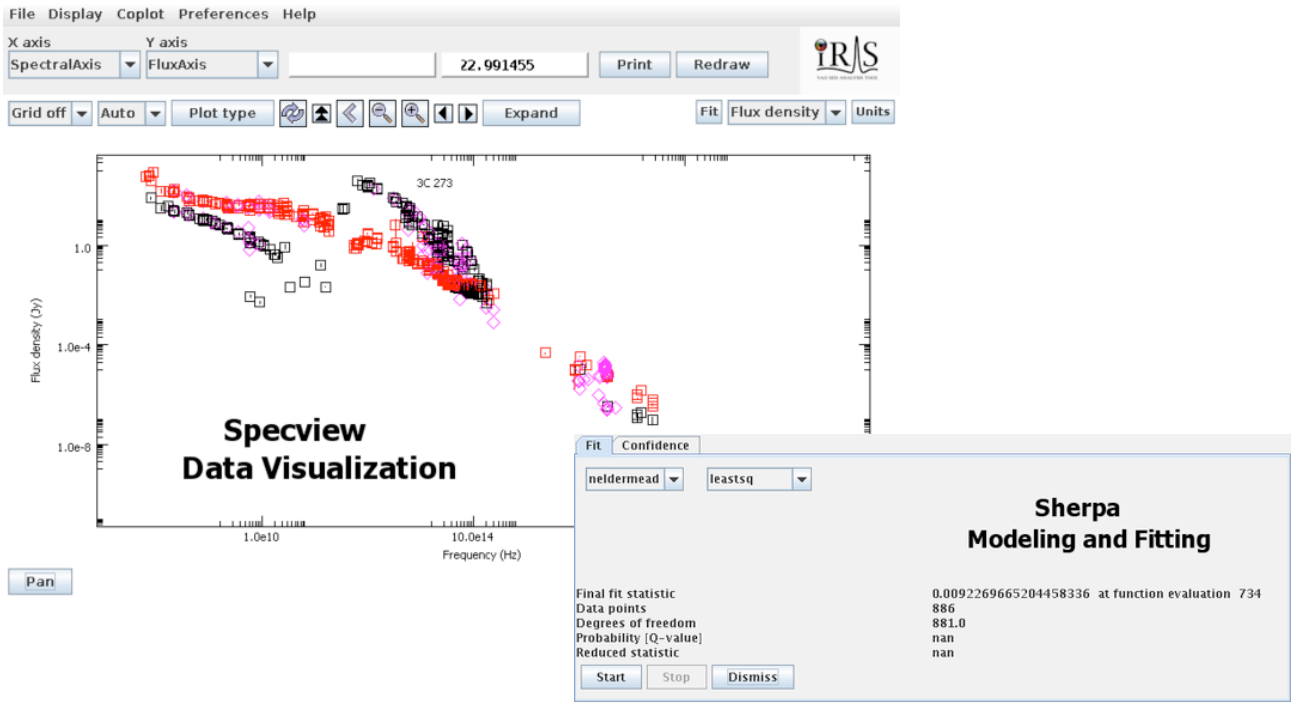

Figure 1. The diagram shows the Iris Data Visualization of SEDs and interaction with the fitting window.

In an example scenario, a user can search and retrieve astronomical data associated with a particular object (e.g., 3C 273) in the VAO Data Discovery Tool; the tool returns a list of data files associated with that object; by highlighting the SED file and choosing "broadcast", the data is sent out via SAMP; Iris responds to data sent via SAMP of type SED and reads in the data. The SED builder in Iris is populated with the SED; upon completion of analysis in Iris the SED is again "broadcast" and picked up by another SAMP aware tool Topcat[7]. In Topcat further visualization capabilities enable the user to continue their analysis of the data. The applications are loosely coupled via the SAMP protocol. Science applications that are SAMP aware can interoperate naturally without file i/o and enable new science analysis customized by the astronomer for their research.

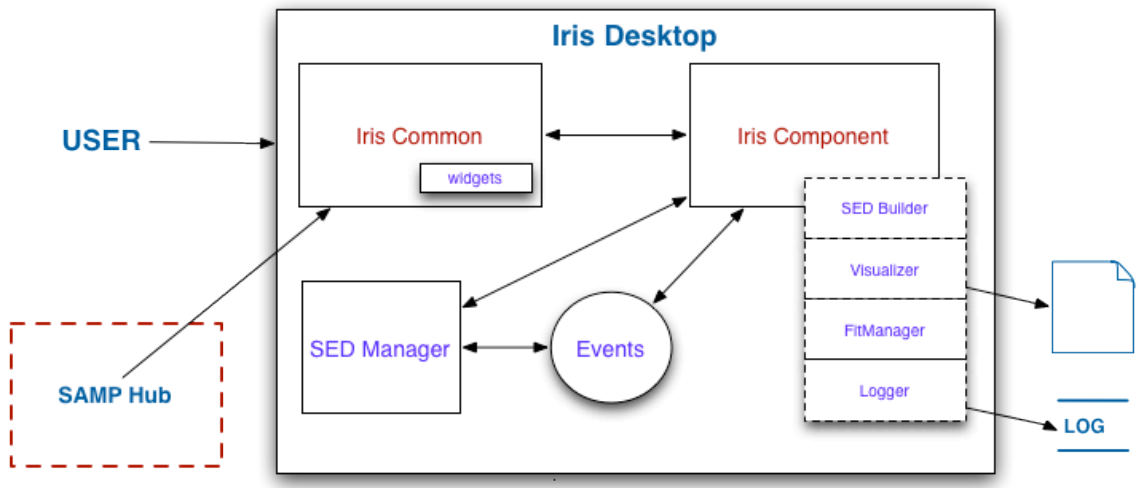

Figure 2. The diagram shows the major components of Iris, and highlights the flow of information through the application. Iris components register with the SED manager and listen for events that they respond to.

# 4. FRAMEWORK FOR DEVELOPMENT

Product development in the VAO follows a lightweight framework that tracks and documents the critical stages of a project. The schedule is driven by higher management decisions to meet project and science priorities and to meet the overall goals of the VAO. Within a project, the team lead manages the product schedule and resources to get the job done within the specified project goals. The key documents used to manage the process are the Project Definition Document, the Product Development Plan, the Project Test Plan, and a Schedule. Internal reviews are scheduled for each of the key stages of development to move forward. Documents are submitted to a *Subversion* (SVN) source code versioning system and repository. The project life cycle and software issues are tracked in the *Jira* system. To release a product, an Engineering Change Request (ECR) documents the software changes and outlines the tests that were performed in preparation for release. The ECR is reviewed and approved by a Change Control Board (CCB) prior to public release.

## 1.1 Product Definition Document

The purpose of the Project Definition Document (PDD) is to summarize the goals of a project and to enumerate the project requirements.

The requirements in the PDD are expressed at a high level. More detailed requirements may exist and can be referenced in the PDD. The PDD includes a header section that identifies the Title, Project ID, Revision number, name of the project and QA&T lead, and Document status (Editing, Review-Ready, Research & Prototyping, Design-Ready). Editing indicates that a document is being edited and not in a final state. Review-Ready indicates that the document is largely finished and waiting for review. Research & Prototyping indicates that the project has been reviewed and approved for the research and prototyping phase. Design-Ready indicates that the project has been reviewed and approved for the Design and Planning phase. A Project Description section describes the goals of the project. The goals are stated in terms of capabilities that the project will enable for users. A section on project interdependencies is outlined to explain the relationship between a project and other projects. The Project Dependencies section identifies how a project will depend on deliverables of another project. Project dependencies can be the underlying standards development work underway in the IVOA of which many of these project depend, or other miscellaneous dependencies that may affect the project schedule. A key component of the PDD is the enumeration of project requirements and identification of how the requirement will be tested. If a science requirements document exists for the project, as in Iris, the enumeration of requirements in the PDD are the specific requirements planned for implementation for a scheduled release. Last, a project roadmap provides a high-level view of the planned development and deployment schedule. The PDD is developed as a *Trac* wiki document and is available without restrictions.

## 1.2 Product Development Plan

The Project Development Plan (PDP) highlights the specific project goals and deliverables in terms of delivered software products or enhancements to existing products. The PDP includes three main components: a description of the design and planned implementation, a project test plan (PTP), and a schedule.

The design plan component summarizes the design of the products the project will produce. It also describes the planned implementation, specifying the programming languages used, and existing tools leveraged. It describes any use of existing IVOA standards as well as dependencies on new IVOA standards to be developed. Detailed designs are maintained in separate documents and linked from the PDP.

The Project Testing Plan (PTP) is a special component that exists as a separable document developed by the QA&T Team (contributed by National Optical Astronomy Observatory) in consultation with the Project Lead (from SAO). The PTP document will continue to be updated as it is carried out through the project lifecycle.

The project schedule enumerates a list of tasks to be assigned to team members and a schedule for completing them. Product documentation is included as a critical part of the schedule.

## 1.3 Project Test Plan

The Project Testing Plan (PTP) is a separate document developed by the QA&T team in consultation with the Project Lead. The PTP document continues to be updated throughout the project lifecycle.

The purpose of the PTP is to describe what tests will be created and verified before products can be released. The PTP in particular must provide a means to test each requirement enumerated in the PDD according to the verification method. The PTP may also describe additional tests that do not specifically validate a PDD-stated requirement.

Testing is accomplished in several stages. Early in the development rounds the internal software integration builds are only made available to the development team so that they can integrate the application and test APIs at the unit level. Later in the process, the software is made available to the informal testers so that initial vetting of the application can be evaluated. Even later, the software undergoes a more coordinated test cycle against the test plan to assure all of the components and platforms have been evaluated as outlined in the PTP. A test lead evaluates and writes up the results. Either further iteration with development or recommendation to release comes from this final stage.

## 5. STAKEHOLDERS

Participation of a stakeholder such as a lead scientist who authors and maintains the requirements specification, handles the technical questions from developers, advises on priorities, and is actively involved as a lead tester is a key element of our process. The science lead represents the science users and brings the view of those users to the development process. They are active participants and part of the development team.

For some applications, a lead scientist may write a vision document as an alternate to a more detailed requirements specification. The vision document captures the needs of the user, the features of the system, and other common requirements for the project in a few pages. It's a sensible choice when behavioral requirements are the primary focus of the application. A requirements specification on the other hand is a more complete description of the software function and algorithms. A clear and thorough understanding of the product to be developed is required when writing requirements.

Another task of the lead scientist is development of science use cases. A use case is simply a paragraph or two of text informally describing what happens during a user analysis session. Use cases are another way to express requirements and they often provide important usability details to the development team. Sometimes simple user interface requirements that are often overlooked when writing the more formal requirements specification or vision document are exposed. Use cases can represent single application instances as well as an entire analysis session describing the expected interaction with other applications. For Iris, use cases were ranked from basic to advanced analysis scenarios. We folded them into our development stages as incremental challenges for the software to meet at each release.

Neither the requirements document nor the use cases are static documents. Both documents need to be maintained as the details of the project evolve.

## 6. SCHEDULE AND RELEASE

Releases are planned as part of the high-level project schedule. Releases usually have "drivers" and often carry along other work at lower priority and completed in the time period since the last release. Depending on the time criticality of the driver, a risk assessment is performed on the lower priority work, and items are included in the release (or not) based on their risk against the schedule.

The schedule is set based on estimates of completion of the driver tasks. When implementation of a project involves cross group dependencies, we also develop a "lien" list that identifies who needs to do what in order for the task to be completed. The lien list is briefed at high-level management meetings so that it is clear at the project level what the dependencies are, and their status, leading to the completion of the project.

Public releases are scheduled at incremental stages of product development over the length of the project. Each stage represents a usable application built from a subset of requirements. Our approach leads to a final product that has real user feedback during project development. In order to reach a release stage, we also develop a series of internal and external beta releases of the application. The beta release is available for unit testing, integration testing, download testing, and science review. The team responds to test feedback from the beta release as it is received. Some problems are fixed immediately while others are logged in the bug tracking system for further review and prioritization. Once the team lead assures that the series of beta release proves to have produced a stable and functional release, more formal testing and review is initiated that leads most often to some iteration with the development team and eventually to a public release. For each release cycle, we revisit the requirements and design phases along with the recorded issues and enhancement requests to gather more details for the next functional set of enhancements.

# 7. COMMUNICATION

Planned communications (e.g., a bi-weekly teleconference) assure issues are resolved quickly and the team is working toward a common vision. We require that status reports be sent to the team lead prior to the meeting. By writing a status report, the developers spend time thinking about their accomplishments and issues in preparation for the meeting. An agenda frequently includes review of noteworthy highlights, an action item review, a schedule review, status reports from each team member, and review of issues. A set of pre-seeded notes with agenda and status reports is assembled by the team lead and sent to the development team prior to the start of the meeting. The pre-seeded notes are annotated by the team lead based on the meeting discussion, and a final set of notes is posted to the team twiki area once the meeting concludes.

A reasonable schedule is another form of communication. Team member participation in the estimates and schedule is an important element for team buy-in. Incremental builds help measure the schedule estimates and allows time for adjustments.

# 8. LESSONS LEARNED

A distributed project like the VAO that is based on part-time efforts of the development group is dependent on efficient time management and dedication of the team members. Developers are assigned approximately 20 to 50% of their time on the project. As managers, we are unaware of the entire set of tasks or responsibilities outside of the VAO effort and schedules that are independent of the VAO work. At times, a developer can spend the allotted time weekly, and at other times, they are tied up with other project deadlines and need to drop out and then make up the time. Due to the variability, good communication and long lead planning is important. For Iris, to provide easy access to project status, a twiki page was developed with Status and Schedule for the group. The team lead manages the content and works to keep current schedule, goals, meeting notes and actions organized in such a way that a simple web page is available to keep the team informed with detailed project information whenever they need it. Much of distributed team management is about organization. With oversight and organization, team members know the project is being managed with the expectation of results. Their time is spent efficiently because the overall plan is organized and has direction. Strong leadership and perseverance of project goals brings a positive response from the team.

A well-defined process ensures that a project is cohesive and stays on track. Project infrastructure needs to respond to team feedback (either directly or through an evaluation of how well the process is practiced) and undergo periodic review. We are revisiting the original VAO project framework and making adjustments to better serve the teams and the project. The participation of a stakeholder such as a lead scientist who manages the technical questions, advises on priorities, and is actively involved as a lead tester is very important. Internal product deliveries with planned test and feedback loops enable the team to focus on incremental steps rather than one big step. Our approach ensures feedback early in the process to assure development is progressing as expected. Beta releases are measured against use cases established early in the process, and they provide the opportunity to assess priorities and make course corrections during development. Finally, frequent scheduled communications (for example a bi-weekly teleconference) assure issues are resolved quickly and the team is working toward a common vision. Finally, easy to access project information ensures the team can pick up quickly if they were side-tracked due to other project responsibilities.

At the VAO, we are developing software to support science analysis across the entire electromagnetic spectrum. We are not serving any one observatory or one specific type of science user. When building tools for "your" data, you can narrow the scope on how data needs to be treated and you can make algorithmic decisions appropriate for your waveband (e.g., X-ray vs. radio) or type of data (e.g., stars vs. galaxies). You know the homogeneous data passing through your application and most often the homogeneous set of users accessing the application. Your users are most often project scientists or frequent users of one of your instruments and data. A small percentage of users are the "outside" user who may depend on helpdesk or colleagues to get through a visit. When you build a tool for the VAO (or more generally the VO), your audience is heterogeneous, likely unknowing of the data or archive, and not interested in the details of VO data model or service standards and protocols. More care needs to be taken in presenting information clearly; you need to design a clean interface to lead users to the intended data or analysis; and you need to work hard to keep your users productive in finding and combining data in ways that are not possible easily with more traditional archives or software.

## 9. CONCLUSION

The software development framework that we describe for the VAO Iris project is similar to the process we follow at the Chandra X-ray Center Data System[8] (CXCDS). Some of the management approaches (e.g. incremental releases, pre-seeded meeting notes, stakeholders) that are described in this paper were adapted from experience with Chandra. The main difference is that at the CXCDS, the development team is co-located, with science teams only a few miles away in another building in Camdridge, MA. The CXCDS team is face-to-face on a daily basis and able to make corrections and adjustments to the work plan, or address problems more easily by meeting. For a distributed team that is also working on a part-time basis, the distance gap needs to be managed with diligence.

The Iris team evolved into a cohesive team during the first few months of the project. The management structure that we describe in this paper was very important in building and sustaining the efforts. More intangible elements that one can only hope is part of a team effort are the willingness of individuals to work collaboratively toward an assigned goal. Folks participated independent of institutional boundaries to build the application. They were willing to contribute in whatever way necessary to get the job done. A good management structure can foster such dedication, but the biggest strength of the Iris team is at the individual level and the willingness of each team member to help.

## 10. ACKNOWLEDGEMENTS

Support for the development of Iris is provided by the Virtual Astronomical Observatory Cooperative Agreement AST0834235 with the National Science Foundation. Support for Sherpa is provided by the National Aeronautics and Space Administration through the Chandra X-ray Center, which is operated by the Smithsonian Astrophysical Observatory for and on behalf of the National Aeronautics and Space Administration contract NAS8- 03060. Support for Specview is provided by the Space Telescope Science Institute, operated by the Association of Universities for Research in Astronomy, Inc., under National Aeronautics and Space Administration contract NAS5-26555. This research has made use of the NASA/IPAC Extragalactic Database (NED) which is operated by the Jet Propulsion Laboratory, California Institute of Technology, under contract with the National Aeronautics and Space Administration.

## REFERENCES

[1] Doe, S., et al, "Iris: The VAO SED Application", arXiv:1205.2419 [astro-ph.IM] (2012).
[2] Berriman, G. B., Hanisch, R. J., Lazio, J. T., Szalay, A., and Fabbiano, G., "The Organization and Management of the Virtual Astronomical Observatory", (these proceedings).
[3] Mazzarella, J. M., Madore, B. F., Helou, G., and the NED Team, "Capabilities of the NASA/IPAC Extragalactic Database in the Era of a Global Virtual Observatory", Proc. SPIE 4477, 20–34 (2001).
[4] Busko, I., "Specview: a Java tool for spectral visualization and model fitting of multi-instrument data," Proc. SPIE 4847, 410–418 (2002).
[5] Freeman, P.F., Doe S., and Siemiginowska A., "Sherpa: a mission-independent data analysis application", Proc. SPIE 4477, 76 (2001).
[6] Taylor, M., Boch, T., Fitzpatrick, M., Allan, A., Fay, J., Paioro T., Taylor, J., and Tody, D., "Simple Application Messaging Protocol Version 1.3", IVOA, <http://www.ivoa.net/Documents/SAMP/> (2012).
[7] Taylor, M., "TOPCAT & STIL: Starlink Table/VOTable Processing Software", ASP Conf. Ser. 347, 29–33 (2005).
[8] Evans, J. D., Evans, I. N., and Fabbiano, G., "The software development process at the Chandra X-ray Center", Proc. SPIE 7019, 701917–701917-9 (2008).